## Astronomy from the Moon in the next decades: From Exoplanets to Cosmology in Visible Light and Beyond

**Jean Schneider [(1)], Pierre Kervella [(2)] and Antoine Labeyrie [(3)]**

*(1) Luth, Paris Observatory, Meudon 92190 France, (2) LESIA, Paris Observatory, Meudon 92190, France (3)  Observatoire de la Côte d'Azur, Nice 0.000,  France*

**Keywords:** solar system - exoplanets -  stellar physics – cosmology - quantum physics

---

**Abstract**

We look at what astronomy from the Moon might be like in the visible over the next few decades. The Moon offers the possibility of installing large telescopes or interferometers with instruments larger than those on orbiting telescopes. We first present examples of ambitious science cases, in particular ideas that cannot be implemented from Earth. After a general review of observational approaches, from photometry to high contrast and high angular resolution imaging, we propose as a first step a 1-metre-class precursor and explore what science can be done with it. We add a proposal to use the Earth-Moon system to test the Quantum Physics theory

## 1. Introduction

Progress on the big questions in astronomy, such as life on certain exoplanets or dark matter, will ultimately require high angular resolution, a large collecting area and access to the full optical spectrum.

The Moon offers unique conditions with its lack of atmosphere, access to the entire spectral band, its position relative to Earth and its low gravity.

More precisely, it combines three advantages:

- its lack of atmosphere allows access to the entire UV/Vis/IR spectrum and avoids atmospheric turbulence

- its low gravity and absence of wind make it possible to install extremely large telescopes with very large instruments, which is impossible for satellites in orbit

- It allows the instruments to be upgraded and to have a very long lifetime, which is impossible for satellites due to their limited amount of fuel.

Large instruments will be installed not before a decade or two. In the meantime instrumental technologies will make in parallel significant progress.

This paper is organised as follows. In Section 2.1 we first describe the observations that are only possible from the Moon, and in particular the ultimate scientific goals of a large telescope or interferometer. In section 2.2 we list some additional benefits of a lunar telescope or interferometer. In section 3 we examine some technical issues for a lunar telescope. In section 4 we discuss instrumental approaches and a 0.3m - 1m class precursor, which will already be able to address important scientific questions. In section 5 we propose another goal for science from the Moon: Earth-Moon quantum correlations.

## 2. Main scientific goals

All astronomy will benefit from the advantages provided by the localisation on the Moon. Here we selected science cases unique from the Moon and a few for domains of high scientific priority.

### 2.1 Scientific cases unique from the Moon

Hereafter, we list a series of science case examples which benefit from the advantages of a lunar based telescope in the UV/Vis/NIR spectral domain. For the medium and far infrared see [1].

#### 2.1.1 Solar system

For the solar system, a lunar telescope can observe toward the Earth and solar system objects.

From the Moon, one can have a global view of the Earth. With an instrument providing only a single pixel image of the Earth, one is in the same conditions as in the imaging of exoplanets. The results of these observations will help to model the spectropolarimetric imaging observations of exoplanets. We go into more details in the section 4.3.1 with the description of the LOUPE instrument.

When a star is on the line going from a lunar telescope to the centre of the Earth, or close to it, the Earth's atmosphere acts as an Earth-sized converging lens [2]. The light rays coming from the star are refracted at different angles depending on their wavelength and the height of their path through the Earth's atmosphere. Some of them will converge exactly on the lunar surface (Figure 1), resulting in a magnification of up to 50,000 |2]. This type of lens is not suitable for imaging, but its magnification can make it useful for rapid photometric observations of optical pulsars for example.

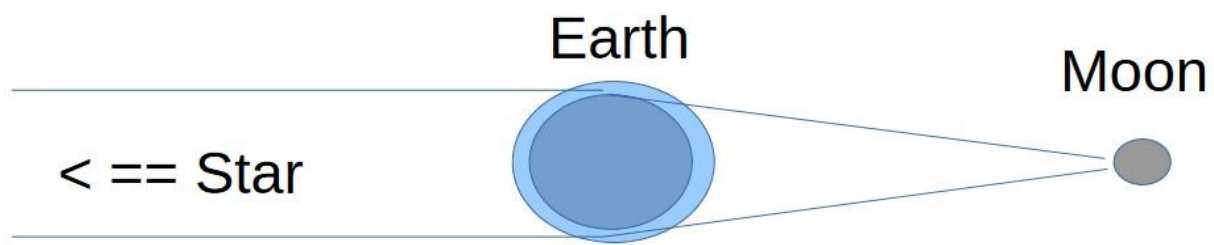

**Figure 1**     Stellar light rays converging  at the Moon surface.

#### 2.1.2  Solar System small bodies

Solar system planets will be better investigated with in situ missions. But, with stellar occultations by solar system bodies from the Moon, one benefits from paralactic effects compared to terrestrial observations. From the Moon one can observe occultations that are not visible from Earth. So one will get more stellar occultations. And compared to ground-based observations, they will add constraints on the shape of the solar system body, rings and orbits.

An unexplored area, the occultation of binary stars by asteroids, is in preparation [3]. A lunar telescope can observe those occultations which are missed from Earth. They will constrain the geometry of unknown orbits, especially for spectroscopic binaries unresolved by imaging.

### 2.2 Exoplanets and exo-moons
### 2.2.1 Very high angular resolution imaging of exoplanet transits

When an exoplanet transits its parent star, it blocks the star's light. Some of the stellar light is seen by transmission through the planet's atmosphere, if there is one. This has led to the proposal to use spectroscopy of this transmitted light to detect molecules in this atmosphere [4]. It was first observed for the planet HD 209458 b, where sodium was detected in the atmosphere [5]. The depth of the transit is $\left(R_{pl}/R_*\right)^2$ where $R_{pl}$ and $R_*$ are the radius of the planet and star. Then the signal-to-noise ratio for a given exposure time and star magnitude is proportional to $R_{pl}/R_*$. . If the spatial resolution $R_\Delta$ provided by a very high angular resolution telescope is smaller than $R_*$ , the signal-to-noise ratio for the same exposure time and star magnitude leads to a gain of $R_*/R_\Delta$ (Figure 2) .

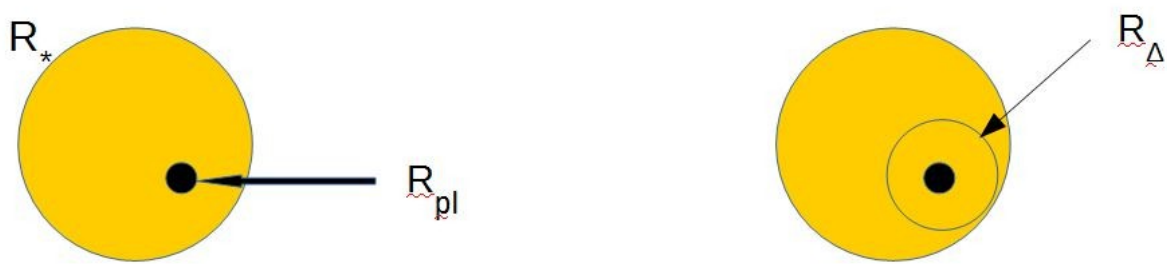

**Figure 2** Very high resolution imaging of an exoplanet transit

When the angular resolution $R_\Delta$ is of the order of $R_{pl}$ some pixels can focus on the planet's atmosphere. They then receive essentially the stellar light, with some molecular absorption lines from the planet atmosphere, leading to a better knowledge of their molecular composition. (Figure 3).

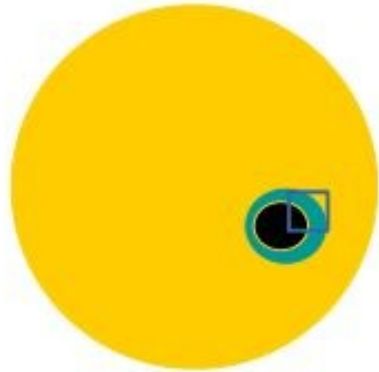

**Figure 3** Imaging of an exoplanet atmosphere transit with $R_\Delta = R_{pl}$. The square symbolizes a telescope pixel

*2.2.2 Imaging of stellar glint on exo-oceans*

Just as terrestrial oceans [6] and Titan lakes [7] reflect the Sun (Figure 4 from |6]), oceans on exoplanets reflect the light of the parent star [8]. It is a direct confirmation of liquid water on the planet. Note nevertheless that glaciers and high-altitude cirrus clouds produce a similar glint, but the latter changes in time with the planet meteorology.

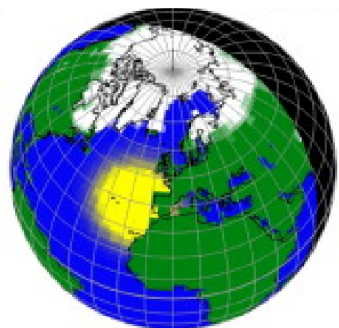

**Fig. 4** Simulation of the Sun glint by terrestrial oceans (Credit Williams & Gaidos 2008)

The planet rotation will modulate the planet specular reflection. This modulation is different from the modulation due to albedo difference from oceans (A < 0.1) and land (A > 0.3). The latter is strictly periodic, with the period of the planet rotation. On the contrary, the occurrence in time of the glint depends on the phase of the planet orbital revolution (Figure 5). In that case, several direct images of the planet must be taken.

There is a contrast difference between glint of oceans and glint of ice (glaciers, cirrus) because the ocean albedo is 0.1 and ice albedo is 0.8. But in both cases the $H_2O$ molecule is involved.

While this kind of observations could be made from the ground, the Moon offers two advantages: the whole spectral range UV/Vis/IR is accessible and the absence of atmosphere improves significantly the photometric precisions.

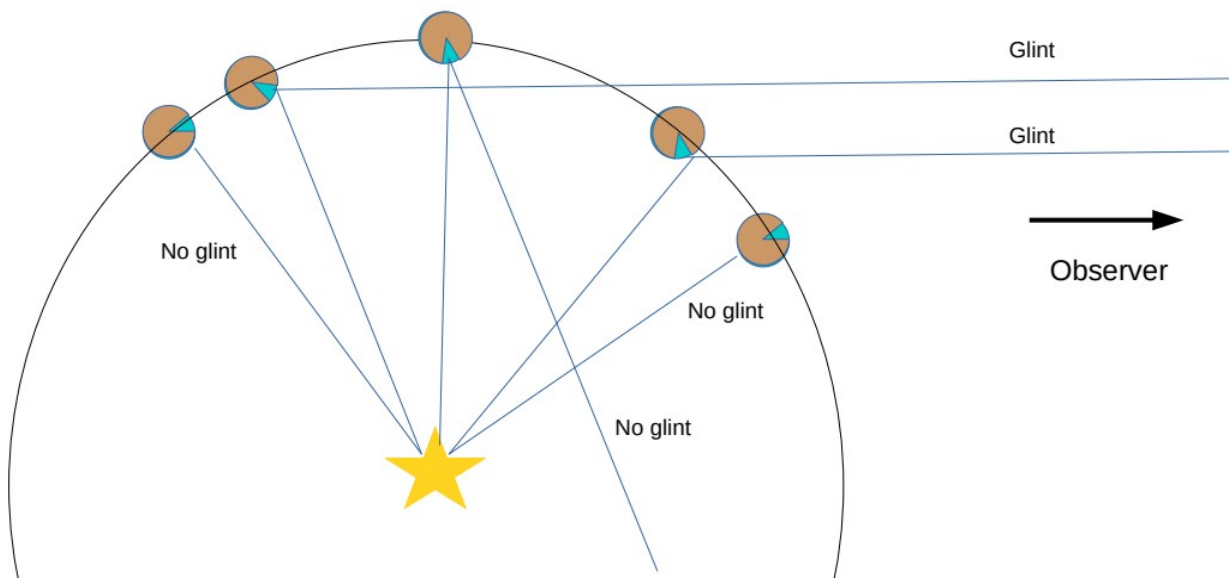

**Figure 5**  Glint observed or not depending on the  planet rotation and orbital phases

### 2.2.3 Detection and internal structure of exo-moons

Moons are very common in the Solar System. Earth-sized moons orbiting exoplanets, or exo-moons, could be detected by direct transit or transit time variation (TTV) with current telescopes [9]. Although more than 5000 exoplanets have been detected to date, the only current exo-moon candidates are controversial or to be confirmed by 30 December 2023 [10]. This may be due to poor SNR due to stellar activity for direct transit or misleading interpretation of transit time variations (TTVs) [11]. Given that the majority of Solar System moons are significantly smaller than Earth, it makes sense to search for small exomoons by direct imaging with an interferometer. An interferometer with a collecting area of 500 m² and a baseline of 200 metres will be able to detect and separate a 0.3 moon at 0.01 au from its parent planet at 5 pc by direct imaging at 1 µ. Spectrographic observation of the moon will detect surface features such as plumes. If it has surface inhomogeneities, a temporal monitoring will reveal the revolution period around its axis. It it is locked to its revolution period around the parent planet, it will put constraints on its internal structure.

*2.3 Extragalactic domain*

*2.3.1. Gravitational lensing of quasars.*

When a background quasar is slightly off the line of sight of a foreground galaxy, its Einstein ring breaks into a larger and a smaller arc [12]. Figure 6 shows the configuration for the double quasar 0957+561 (mV = 17). Images with an angular resolution at least 0.1 mas will measure the length and width of these arcs, and their curvature will constrain the shape of the lensing galaxy. Given the known quasar and lense distances and the observed separation of the two quasar images of 6 arcsec [13], for a spherical lensing galaxy (B) and a spherical source with a radius of 100 pc, the length and thickness of the two source images are predicted to be 24 and 16 mas and 0.2 and 0.3 mas respectively. Deviations from these predictions will constrain the shape of the lens and the source

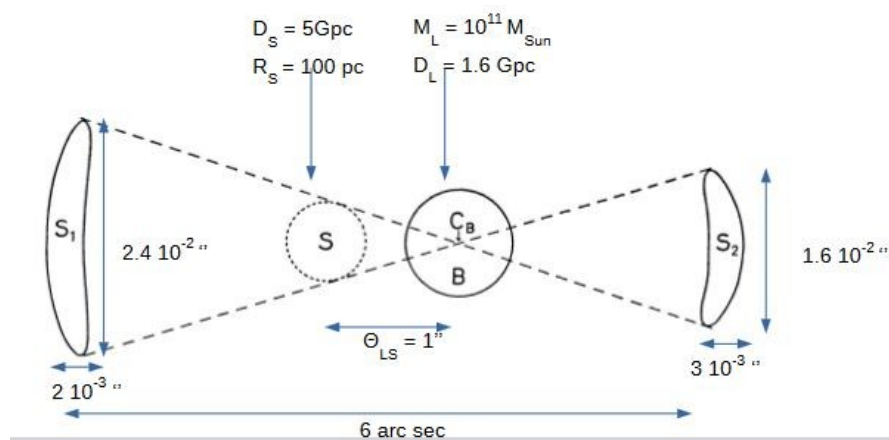

**Figure 6** Geometry of a lensed quasar, adapted from [14] for the double quasar QSO 0957+561. $R_S$ and $D_S$ are the radius and the distance of the quasar source, $M_L$ and $D_L$ are the mass and the distance of the lensing galaxy. The figure is not on scale, and the lens galaxy is supposed to be spherical.

*2.3.2 Dark matter distribution*

Einstein rings can be disrupted by anomalies in dark matter distribution in the case where dark matter is made of axions, beyond the standard particle model [15] (Figure 7).

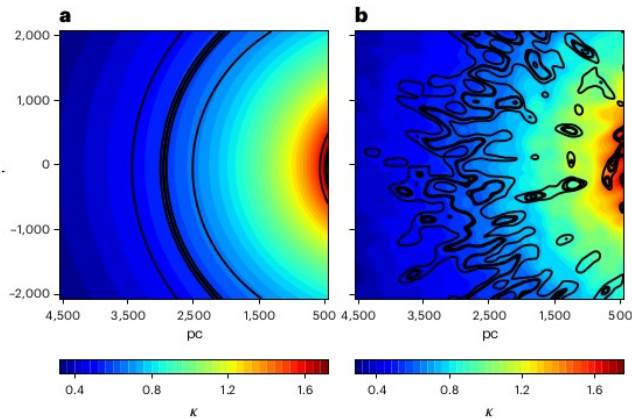

**Figure 7 Left:** Einstein ring in case of dark matter distributed smoothly in Weakly Interacting Massive ParticleS (WIMPS). **Right:** Einstein ring modulated by fluctuations in dark matter distribution ( [15]).

**2.2 Additional benefits from large lunar instruments**

Access to the entire UV/Vis/IR spectral range with large to very large telescopes will allow much better characterisation of many celestial bodies. For exoplanets, for example, it will provide new molecules, the surface colours of rocky planets and their temporal variations due to their rotation, volcanoes or seasons. Precise tracking of their orbits will help to determine the gravitational interaction between planets and thus constrain their masses. It will provide very precise architectures of galaxies and detect very faint galaxies and give their organisation better than Euclid. It will reveal the detailed environment of supermassive black holes.

**3 Issues**

*3.1 Dust*

There will be two main sources of lunar dust, meteorite impacts and human activity. They will mainly affect the reflectivity or transparency of optical surfaces. Several projects aim to measure its properties, e.g. [16]. It is therefore important to take countermeasures to eliminate its effects on instrumentation. For example, electromagnetic removal has been proposed [17]. In fact, the 15 cm telescope operations of the Chang'E 3 mission have shown that the dust problem may not be too severe [18].

*3.2 Meteorids*

Meteoroids have not affected the Cheng'e 3 mission for more than 18 Month [18]. But future lunar instruments will remain on the Moon for decades. For such long time scales, meteoroids can present a threat to these instruments. If finally selected, the ESA LUMIO CubeSat mission [19] (currently in Phase B development [20]) will provide statistics of the meteorids reaching the lunar surface. Measures to prevent them should be investigated.

*3.3 Lunar seismology*

As analysed in [21], the seismic activity of the Moon can be an issue for lunar interferometers. However, as the authors note, "a lunar-based version could be considered in the long term if a human presence would allow maintenance and upgrades, leading to a longer lifetime with continuous performance improvement". Indeed, in the long term, humans will be able to add acoustic filters to the interferometers, similar to what is done with terrestrial gravitational wave detectors.

*3.4 Regolith*

The mechanical properties of the regolith at a telescope site need to be investigated prior to installation or consolidated to ensure stability.

*3.5 Day-night thermal shock*

The temperature difference between day and night on the Moon is 350 K. When a given location crosses the terminator during the lunar rotation, the temperature jump occurs in a few hours, threatening the electronics and materials of instruments. For example, India's Chandrayaan-3 lunar probe was damaged by the brutal temperature change. It was therefore necessary to add radiation sources to mitigate the temperature jump. A preliminary study of lunar mission night survival is needed [22].

*3.6 Robotisation or human assistance ?*

Some complex operations may need human assistance, like in the ISS. One cannot predict a priori which ones cannot be robotized. Preliminary ground-based experiments will clarify the problem. Mounting of delicate instrumentation on the Moon by human fingers can only be made inside lunar habitation-like laboratories.

*3.7 Location*

The best location for a lunar telescope depends on two factors: the physical conditions of the site (temperature, soil quality, solar illumination) and its scientific objectives. For example, telescopes pointing towards Earth must be placed on the near side of the Moon. From the point of view of target observability, they can be placed almost anywhere. At the lunar poles, only half the sky is visible, but all the time. At the lunar equator, the whole sky is visible, but only half the time. For observations of or towards the Earth, the optimum is not far from the lunar equator.

*3.8 Cost*

Although cost is an important issue, it is beyond the scope of this paper and should be estimated for each aspect by a specialised department.

**4 Instrumental approach**

Installation of large lunar telescopes has already been proposed by several authors (e.g. [23], [24]).

### *4.1 Future large instruments*

A priori, any telescope can do spectro-photometry and spectro-polarimetry. For example, the 620 m2 array of 6.5 m telescopes at the lunar poles (Life Finder Telescope At Lunar Poles LFTALP) is an array dedicated to transits [25].

Telescopes with a prime mirror larger than 30 m will detect objects fainter than mV = 26 in visible light with a bandwidth 100 nm in a 1 hour exposure with a SNR 5, suited for faint free-floating planets and very distant galaxies or quasars.

### *4.2 Very high angular resolution*

#### *4.2.1 Standard interferometer*

A standard interferometer project, similar in its beam combination principle to the European VLT Interferometer (VLTI) with GRAVITY science instruments [26], has recently been proposed for the Moon [27]. It could have a 10 km baseline, leading to a 10 µas angular resolution at 0.5 micron.

The Moon rotates much more slowly than the Earth. This results in a much shorter continuous variation of the optical delay for a given duration of an observation. For a 3 hour observation, only one metre of delay would be sufficient, compared to 30 metres on Earth. However, in order to have sufficient coverage of the sky, it is still necessary to compensate for the potentially large fixed optical delay required to point to a given position in the sky. For this purpose, we propose to use a set of optical fibres of different lengths, selected by a mechanical or opto-electronic beam-switching device. The principle would be the same as the two-stage system that has been operating successfully at the CHARA array for more than 20 years [28], [29]. A set of fixed length incremental delay lines would be associated with each telescope of the lunar interferometer array. They could be positioned either in each telescope array or alternatively in the central beam combination system.

On Earth, the angular resolution of interferometers is limited to a few mas [30]. That limitation comes from the atmospheric turbulence. The absence of atmosphere on the Moon, combined with a precise control of delay lines, allows an angular resolution λ/D = 10 µas for a 1 km baseline D at 0.5 µ, sufficient to explore the transit by a 1 $R_{Earth}$ planet atmosphere at 10 pc, as shown in Figure 4.

Quantum-enhanced methods for pushing to extremely large base lines ([31], [32]) requiring sophisticated quantum technologies not adaptable to free-floating telescopes, can be installed in the future on the Moon. A sub-microarcsecond angular resolution is now in view [31].

#### *4.2.3 Hypertelescope*

A hypertelescope is a 2D array of many sub-pupil mirrors, with a 'pupil densifier' at the recombined focus, providing direct 2D images instead of complicated fringe images for extended sources [23] . The idea of the pupil densifier is to dilate the images behind the Fizeau focus with an optical setup to combine them into a simple connected pupil, leading to a standard 2D image. (Figure 8).

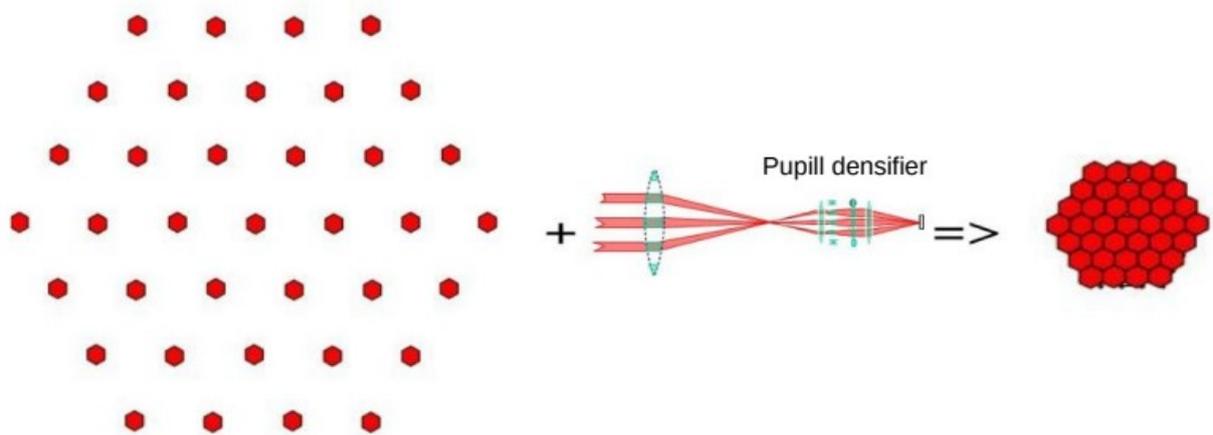

**Figure 8**. The principle of pupil densification

For a hypertelescope, the final image will have N = D/s pixels, where D is the hypertelescope baseline and s is the separation between the sub-apertures, greater than the diameter d of each sub-aperture. The concept has been proven, on a very small scale, with 78 sub-apertures in the laboratory and in the sky [33]. On the Moon, the hypertelescope can be implemented in two configurations: an array of apertures on the lunar surfaces with delay lines to compensate for the Moon's rotation, like a standard interferometer, or an array of fixed mirrors on a paraboloid installed in a lunar crater (Figure 4).

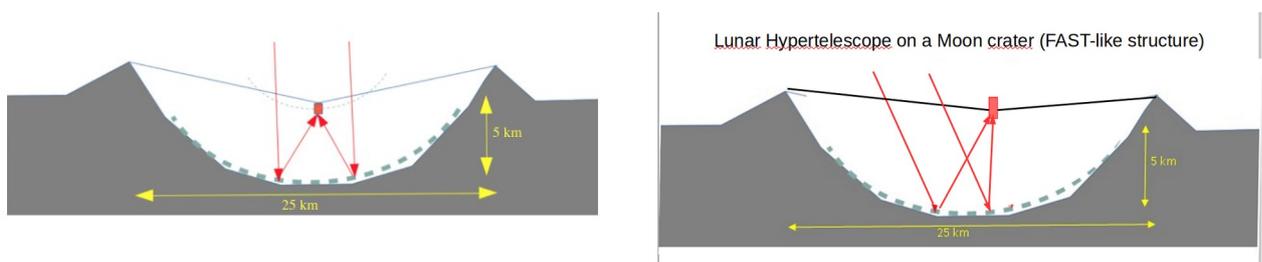

**Figure 4** Sketch at approximate scale of a Crater-Nested Lunar Hypertelescope with its focal receiver suspended from cables and movable along the focal surface (fine dotted arc). The locus of the mirror array (fat dotted arc) is either paraboloidal actively deformable by applying small tip-tilt piston corrections on its segments, or fixed and spherical if a corrector of spherical aberration is added in the focal optics. No delay lines are needed but the "meta-mirror" size is limited to about 20 km by the maximal 6 km depth of impact craters for an effective meta-aperture of 5–10 km. The mirror elements are either carried by separate fixed tripods or by a hammock-like cable netting. The figure at right shows how the focal instruments can be moved to follow the target.

### 4.3 High contrast imaging

Some objectives require not only an angular resolution of 0.1 mas and a sensitivity to $m_V = 26$, but also a high rejection factor for a faint object close to a bright one. It is for example the case of a planet reflecting the light of its parent star. For the direct imaging of exoplanets, a high rejection factor of the parent star is required. For a planet radius 1 $R_{Earth}$ and albedo of 0.3 and star-planet distance of 1 AU, the planet to star brightness ratio is $10^{-10}$ .

To achieve this goal, there are three approaches: a coronagraph in the focal plane, an self-extinction of the star by nulling and an external occulter in orbit. An external occultor in orbit requires a position accuracy of   +/- 1 meter [34] which would require propellant consumption [35]. Nulling systems have achieved a rejection factor of only $10^5$ [36]. By comparison, coronagraphs seem more promising. Since the classical Lyot coronagraph, these devices are making constant progress. For instance, a rejection factor $10^7$ has been obtained with a vector coronagraph [37] and a rejection $10^{10}$ is in view with this new approach [38]. An additional wavefront sensor in focal plane gives an additional rejection factor of 10 [39].

### 4.3  A precursor

With the ultimate science goals above in mind, a 30 cm - 1 m class telescope will already explore significant science cases.

#### 4.3.1 Solar system

Even before a 30 cm telescope, LOUPE (aperture 2 cm, 1 litre volume [40]) will observe the Earth in a singe pixel from multiple phase angles, and its variations over the seasons and the reflection of the Sun by the oceans [41].  Exoplanets will initially be seen only as a point, just as LOUPE will see the Earth. The comparison of LOUPE results with images of the Earth from a 30 cm - 1m telescope will validate the conclusions drawn from a single pixel image and applied to single pixel observations of exoplanets.

Standard stellar occultations of solar system bodies are routinely performed on Earth with 30 - 50 cm class telescopes, and are therefore not difficult to perform on the Moon, and will detect occultations not visible from Earth.

*4.3.2 Exoplanets*

Exoplanet transit spectroscopy is routinely performed on Earth and in space with 1-m class telescopes. Lunar observations will open up the full UV/Vis/IR spectral range and discover new molecules.

Exoplanet detection by microlensing is routinely performed on Earth and in space by 1-m class telescopes. Lunar observations will provide additional paralactic constraints by comparison with observations from the ground and the Euclid and Roman space telescopes.

*4.3.3 Extragalactic*

The LOUVE project (Lunar Optical UV Explorer - [42]) is a 30 cm project to take advantage of the absence of lunar atmosphere to perform spectrophotometry of bright sources in the UV.

In the absence of atmosphere, photometric precision is better than on Earth, and a 1 metre telescope can detect point sources at the 5 sigma level down to mV = 17 in 10 minutes. For quasars, their variability provides clues to understanding the accretion process towards supermassive black holes. For lensed quasars, their variability can be used to measure the time delay between the multiple images, providing a measurement of the Hubble constant [13].

**5 Quantum Physics**

In addition to telescopes, the Moon also offers the opportunity to test one of the foundations ofmulti astrophysics that is still under debate: the observational theory of quantum mechanics.multi

The problem in quantum mechanical theory of observation is as follows. The quantum entanglement between pairs of photons emitted by a spin 0 source (by the statistical correlations of their polarisation) can theoretically go to infinity. It has been proposed to test this statement for Earth-Moon distances ([43], [44]): the two photon detectors would be placed one on the Earth and the other on the Moon. Recently, a variant implementation has been proposed [45]. The source of the photon pairs would be at the Earth-Sun L4 or L5 Lagrange point, with one detector on the Earth and the other on the Moon (Figure 5). Actually, both experiments should be performed, since, given our ignorance of what might be beyond standard quantum mechanics and the unclear relation between Quantum Physics and classical space, one does not know whether a variation of entanglement with distance depends on the length of the photon paths or on the distance between the detectors. Chronometric perturbations by the Earth atmosphere [45] can be avoided and geometric attenuation can be lowered by putting an intense source on the Loon and the two detectors at L4 and L5.

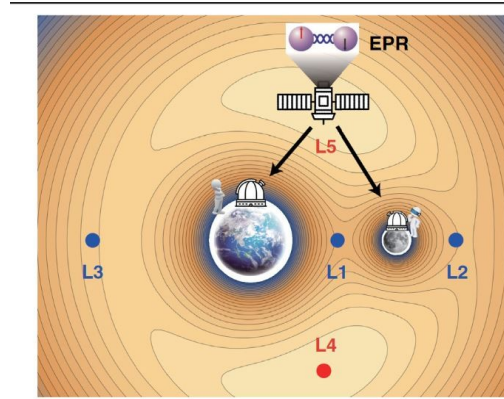

**Figure 10**  A source of entangled photons pairs at the Earth-Sun L4 or L5 Lagrange point and one detector on the Earth and the other on the Moon  [48].

One may ask whether there can be a priori theoretical estimates for a distance scale D or a propagation velocity V of non-standard correlations. Playing only with the usual fundamental constants h, c and G can only give V = c or V = ∞ and D = Planck length $10^{-33}$ cm, which is excluded. Playing with less fundamental constants like the quark mass Mquark allows to multiplymulti these values by any power N of the dimensionless constant $GM^2_{quark}$ /hc ~ $10^{-39}$. For small valuesmulti of N, such as -1 or +1, one obtains V = $10^{+/-39}$ c and D = $10^{-33}$ cm, which are excluded, or $10^{39}$ cm, which is unobservable by a lunar experiment. Another source of a priori estimates for the distance scale D could come from the MOND theory as an alternative theory to dark matter (D ~ 10 kpc, [46]) or the distance D = cT = 100 kpc derived from the time scale T = $10^8$ yr for the "spontaneous collapse" theory [47]. As can be seen, the range of possible a priori predictions has no firm constraint, and only experiment may eventually provide a constraint. None of these estimations leads to a distance up to about 10 times the Earth-Moon distance, which is the only scale for which a deviation from the quantum mechanical prediction is observable. The detection of a deviation from the quantum mechanical prediction would therefore lead to a new fundamental constant.

**References**
1.  Maillard J.-P.-P. 2023 Infrared astronomy beyond JWST: the Moon perspective. This issue
2. Kipping D. 2019. The "Terrascope": On the Possibility of Using the Earth as an Atmospheric Lens *PASP* **131**, 114503 doi: <u>10.1088/1538-3873/ab33c0</u>

3. Souami D. 2023 Private communication
final-PTA_Moon-15February-Schneider-RSTA20230071.docx

4. Schneider J. 1994 On the search for O2 in extrasolar planets  Astr. & Spa. Sci. , **212** , 321-325 doi:10.1007/BF00984535

5. Charbonneau D., Brown T., Noyes R. & Gilliland R. 2002 Detection of an Extrasolar Planet Atmosphere ApJ, **568**, 377-384 doi:10.1086/338770multi

6. Williams D, Gaidos E. 2008 Detecting the glint of starlight on the oceans of distant planets. Icarus **195**, 927–937. (doi:10.1016/j.icarus.2008.01.002)

7. Stephan K., Jaumann R., Brown R. et al. 2010 Specular reflection on Titan: Liquids in Kraken Mare *Geophys. Res. Lett.* **37,** L7104  doi: 10.1029/2009GL042312

8. McCullough 2006. Models of Polarized Light from Oceans and Atmospheres of Earth-Like Extrasolar Planets.  *ApJ*, submitted doi: 10.48550/arXiv.astro-ph/0610518

9. Sartoretti P. & Schneider J. 1999 On the detection of satellites of extrasolar planets with the method of transits *Astron. & Astrophys.* **134**, 553-560 doi:10.1051/aas/1999148

10. "Drake equation" of exomoons -- a cascade of formation, stability and detection. *Universe*, submitted doi:10.48550/arXiv.2311.05390

11. Yahalomi D., Kipping D., Nesvory D. et al. 2023 Not-so-fast Kepler-1513: a perturbing planetary interloper in the exomoon corridor *MNRAS*, **527**, 620-639 doi:10.1093/mnras/stad3070

12. Cassan A. 2023 Interferometric visibility of single-lens models: The thin-arcs approximation *Astron. & Astrophys*. **676** A110 https://doi.org/10.1051/0004-6361/202142429

13. Vanderriest Ch., Schneider J., Herpe G. et al. 1989, The value of the time delay delta T (A,B) for the 'double' quasar 0957+561 from optical photometric monitoring. A*stron. & Astrophys.* **215,** 1- 13 https://ui.adsabs.harvard.edu/abs/1989A%26A...215....1V/abstract

14. Refsdal S. 1964. The gravitational lens effect. *MNRAS*, **128**, 295-306
 doi:10.1093/mnras/128.4.295

15. Alfred A. et al. 2023 Einstein rings modulated by wavelike dark matter from anomalies in gravitationally lensed images. *Nature Astronomy*  **7**, 736–747  doi:10.1038/s41550-023-01943-9

16. Lolachi et R., Glenar D. and Stubbs T. 2023  Optical monitoring of the dust environment at lunar surface exploration sites *Acta Astronaut.* **234,** 105709 doi:10.1016/j.pss.2023.105709

17. Hirabayashi M., Hartzell C., Bellan P., Bodewits D., Delzano G. et al. 2023 Electrostatic dust remediation for future exploration of the Moon *Acta Astronautica* **207**, 392-402 doi:10.1016/j.actaastro.2023.03.005

18. Wang J., Meng et al. 2015 18-Months operation of Lunar-based Ultraviolet Telescope: a highly stable photometric performance  *Astrophys Space Sci* **360**, 10 doi:10.1007/s10509-015-2521-2

19. Topputo F., Merisio G., Franzese V. et al. 2023. Meteoroids detection with the LUMIO lunar CubeSat. *Icarus* **389**, 115213, doi:1016/j.icarus.2022.115213

20. Ferrari F., Topputo F., Merisio G. et al 2023 The LUMIO CubeSat: Detecting Meteoroid Impacts on the Lunar Farside. 54th Lunar and Planetary Science Conference, held 13-17 March, 2023 at The Woodlands, Texas and virtually. LPI Contribution No. 2806, id.1456

https://www.hou.usra.edu/meetings/lpsc2023/pdf/1456.pdf

21. Bely P.-Y., Laurance R., Volonte S. et al. 1996 Kilometric baseline space interferometry *SPIE* **International Symposium on Optical Science, Engineering, and Instrumentation, Denver, CO, United States** doi:10.1117/12.255123

22. Biswas J., 2023 Rerating of Electronic Components for Improved Lunar Mission Night Survival. Doktors der Ingenieurwissenschaften Dissertation Technische Universität München TUM School of Engineering and Design
https://mediatum.ub.tum.de/doc/1687567/mhrhcbkk3lddt5530dakus7gi.Janos%20Daniel%20Biswas.pdf

23. Labeyrie A. 2021. Lunar optical interferometry and hypertelescope for direct imaging at high resolution. *Phil. Trans. Roy. Soc.* **379,** 2019057
doi:10.1098/rsta.2019.0570

24. Flahaut J., van der Bogert C., Crawford I. & Vincent-Bonneau S. 2023. Scientific perspectives on lunar exploration in Europe. *Nature Micrograv.* **9**, 50. doi:10.1038/s41526-023-00298-9

25. Angel R., 2023. A 600 m² array of 6.5 m telescopes at the lunar pole. Phil. Trans. Roy. Soc. This issue

26. Gravity+ Collaboration 2022 The GRAVITY+ Project: Towards All-sky, Faint-Science, High-Contrast Near-Infrared Interferometry at the VLTI *The Messenger* **189,** 17-22 doi: 10.18727/0722-6691/5285

27. Kervella P. 2023 Optical interferometry from the Moon CNES Workshop Lune https://hal.science/hal-04090335

28. Ten Brummelaar et al. 2005, First Results from the CHARA Array. II. A Description of the Instrument *ApJ* **628**, 453 doi:10.1086/430729

29. Setterholm et al. 2023 MYSTIC: a high angular resolution K-band imager at CHARA JATIS 9, 025006 doi:10.1117/1.JATIS.9.2.025006

30. Martinod M.-A., Defrère D., Ireland M. et al. 2023. High-angular resolution and high contrast multi observations from Y to L band at the Very Large Telescope Interferometer with the Asgard Instrumental suite. *JATIS,* **9**, 025007 doi:10.1117/1.JATIS.9.2.025007

31. Bland-Hawthorn J., Sellars L. & Bartholomew G. 2021. *J. Opt. Soc. Amer. B.,* **38**, A86-A98 doi:10.1364.JOSAB.42.4651

32. Dravins D. Barbieri C., Fosbury E. et al. 2006. *QuantEYE* : The Quantum Optics Instrument for OWL. In *Proc. IAUS* **232**, 506-507 doi:1017/S1743921306001220

33. Gillet S. et al. 2003 Imaging capabilities of hypertelescopes with a pair of micro-lens arrays *Astron. & Astrophys.* **400**, 393-396 doi:10.1051/0004-6361:20021686

34. Chen A., Harness A. & Melchior P. 2022. Lightweight starshade position sensing with convolutional neural networks and simulation-based inference. *JATIS.* **9**, 2.025002 doi:10.117/1.JATIS.9.2.025002

35. Hatzes A. & Liseau R. 2018. Future Exoplanet Space Missions: Spectroscopy and Coronographic Imaging. In In: Deeg, H., Belmonte, J. (eds) Handbook of Exoplanets . Springer, Cham. doi:10.1007/978-3-319-55333-7_88

36. Laugier R., Defrère D., Woillez J. et al. 2023 Asgard/NOTT: L-band nulling interferometry at the VLTI. I. Simulating the expected high-contrast performance. *Astron. & Astrophys.* **671**, L110 doi:10.1051/0004-6361/202244351

37. Doelman D., Ouellet M., Potier A. et al. 2023 Laboratory demonstration of the triple-grating vector vortex coronagraph. In *SPIE Techniques and Instrumentation for Detection of Exoplanets XI; San Diego, California, United StatesConference Series.* **12680**, 126802C doi:10.1117/12.2677432

38. Doelman D., Por E., Ruane G., Escuti M. and Snik F. 2020. Minimizing the Polarization Leakage of Geometric-phase Coronagraphs with Multiple Grating Pattern Combinations. *PASP* **132** 045002 doi:10.1088/1538-3873/ab755f

39. Lin J., Fitzgerald M., Xin Y. et al. 2023. Real-time Experimental Demonstrations of a Photonic Lantern Wave-front Sensor. *ApJ. Letters* **959**, L34 10.3847/2041-8213/ad12a4 ]

40. Klindzic D., Stam D., Snik F. et al. 2020. LOUPE: observing Earth from the Moon to prepare for detecting life on Earth-like exoplanets *Phil. Trans. Roy. Soc. A.* **379**, 20190577 doi: 10.1098/rsta.2019.0577

41. Trees V. & Stam D., 2019 Blue, white, and red ocean planets - Simulations of orbital variations in flux and polarization colors. Astron. & Astrophys. **626**, A129 doi:10.1051/0004-6361/201935399

42. Ehrenreich D. 2023. A telescope on the Moon: the Lunar Optical and UltraViolet Explorer (LOUVE). In the ESA Campaign Ideas for exploring the Moon with a large European lander. Available at
https://ideas.esa.int/servlet/hype/IMT?documentTableId=45087150958286643&userAction=Browse&templateName=&documentId=e8690030269e031e292b6bb05791f48d

43. Schneider J. 2009. Constraining the Correlation Distance in Quantum Measurements White paper submitted to the ESA Fundamental Physics Roadmap Advisory Team doi:10.48550/arXiv.1001.2135

44. Schneider J., 2021 Quantum correlations at earth-moon distance, *Exp. Astron.*, **51**,1 767-1772 doi:10.1007/s10686-021-09751-7

45. Cao et al. Bell Test over Extremely High-Loss Channels: Towards Distributing Entangled Photon Pairs between Earth and the Moon Phys. Rev. Lett. **120**, 140405 doi:10.1103/PhysRevLett.120.140405

46. Soussa M. and Woodward R. 2004. A generic problem with purely metric formulations of MOND. *Phys. Lett. B.* **578**, 253-258 doi:10.1016/j.physletb.2003.10.090

47. Ghirardi G., Rimini A. and Weber T. 1986 Unified dynamics for microscopic and macroscopic systems. *Phys. Rev. D* **24**, 470-491 (doi:10.1103/PhysRevD.34.470 )

**Acknowledgments**
The authors are grateful to two anonymous referees for their constructive comments. JS is grateful to Jean Surdej for discussions and to Darren William, Yuan Cao and Amruth Alfred for their permissions to use their figures.